\def \kpc {\,\rm kpc}
\def \kms {\,\rm km~s^{-1}}
\documentclass{aa}
\usepackage[]{times}
\usepackage[fleqn]{amsmath}
\usepackage{epsfig}
\begin{document}

   \thesaurus{04  (
               {08.11.1 
                10.06.2 
                10.11.1 
                10.19.1 
                10.19.3 
                12.04.1 })}

    \title{The   local    stellar  velocity    distribution     of  the
           Galaxy\thanks{Based on data   from the Hipparcos  astrometry
           satellite}}
    \subtitle{Galactic structure and potential}
    \author{O. Bienaym\'e \inst{}}

     \institute{CDS,  Observatoire Astronomique de Strasbourg,
     11 rue de l'Universit\'e, F-67000 Strasbourg, France}

     \offprints{O. Bienaym\'e}

     \date{Received / accepted}

    \maketitle
   \begin{abstract}

 The velocity distribution of neighbouring  stars is deduced from  the
 Hipparcos  proper motions.  We  have  used a  classical Schwarzschild
 decomposition  and    also    developed   a  dynamical    model   for
 quasi-exponential  stellar discs.  This model  is a 3-D derivation of
 Shu's  model  in  the framework of  St\"ackel   potentials with three
 integrals of motion.

 We determine the solar motion relative  to the local standard of rest
 (LSR) ($U_\odot=9.7\pm0.3\kms$,         $V_\odot=5.2\pm1.0\kms$   and
 $W_\odot=6.7\pm0.2\kms$), the density and kinematic radial gradients,
 as well as the local  slope of the  velocity curve.  We find out that
 the  scale  density  length of   the  Galaxy is  $1.8\pm0.2\kpc$.  We
 measure  a  large  kinematic scale  length   for  blue (young) stars,
 $R_{\sigma_r}=17\pm4\kpc$, while for red stars (predominantly old) we
 find                   $R_{\sigma_r}=9.7\pm0.8\kpc$               (or
 $R_{\sigma_r^2}=4.8\pm0.4\kpc$).

 From the  stellar disc dynamical model,   we determine explicitly the
 link   between the   tangential-vertical  velocity $(v_\theta,  v_z)$
 coupling  and the local shape  of the  potential.  Using a restricted
 sample of 3-D velocity   data, we measure $z_o$,   the focus of   the
 spheroidal  coordinate  system  defining  the  best fitted  St\"ackel
 potential.   The  parameter  $z_o$ is  related  to   the tilt of  the
 velocity ellipsoid and more fundamentally to the mass gradient in the
 galactic disc.  This parameter is  found to be $5.7\pm1.4\kpc$.  This
 implies that the galactic potential is not extremly flat and that the
 dark matter component is not confined in the galactic plane.

    \keywords{ Stars:  kinematics -- Galaxy: fundamental parameters --
 Galaxy: kinematics and dynamics  --  solar neighbourhood --   Galaxy:
 structure -- dark matter }

    \end{abstract}

 \section{Introduction}
 The Hipparcos proper motion measurements (\cite{Hip}) of neighbouring
 stars radically enlarge the size  and quality of kinematical unbiased
 samples and  provide   a  new  opportunity  to  reconsider  the local
 galactic structure and solar neighbourhood kinematics.  Visualization
 and   analysis  of  the local  velocity   field   from the  Hipparcos
 tangential velocities  (Chereul et al, \cite{ccb97}, \cite{ccb98}) or
 combined with radial  velocities  (Figueras et al,  \cite{fg97}) show
 the large structuring of phase space and  reveal the process of phase
 mixing with time.  Clearly a much better  understanding of  the local
 kinematics and of the galactic  dynamics  will be obtained while  the
 real complexity of the local kinematics is now apparent.

 In this paper,  we  re-examine the  classical kinematic analysis.   We
 focuss  this study on the  two following points:
 \linebreak
 1) it is   necessary to build   a dynamically coherent model of   the
 kinematics  since an analysis  based on  the Jeans  equations is  not
 sufficient and a more  rigorous approach follows from using solutions
 of  the Boltzmann   equation;  2) we measure  the  galactic gradients
 (stellar  density and kinematics, slope of   velocity curve) from the
 local kinematic data and constrain the potential shape and the radial
 forces close to the plane.  We  deduce the flattening of the galactic
 potential   (and    mass distribution) in   the   Galaxy.   This is a
 complementary approach to the more classical $K_z$ problem, i.e.  the
 analysis of the force perpendicular  to the galactic plane (Cr\'ez\'e
 et al, \cite{cc98}).

 In  this paper, we analyse  the Hipparcos proper motions, first using
 only a Schwarzschild  decomposition (\S2) then  with the help of a 3D
 dynamical galactic  model (\S3)  which is  a generalization  of Shu's
 model.  Conclusions regarding the solar motion  relative to the local
 standard  of rest  (LSR),  density and  kinematic gradients are drawn
 (\S4).  We finally present  a  first observational constraint  on the
 radial potential bending at the solar position (\S5).

 \section{Kinematic data}
 The Hipparcos  data allow us  to  build a kinematic  sample free from
 kinematic bias, that is the properties used to  define the sample are
 not a kinematic parameter or  do not explicitly depend on velocities.
 Survey stars, labeled S on  column H68 in HIP catalogue (\cite{Hip}),
 satisfy this  criteria.  They define a  magnitude limited sample from
 HIC  (\cite{Hic})  magnitudes with   dependence on galactic latitude,
 $V\le  A+1.1 \sin  |b|$, with $A=7.9$  for spectral  types earlier or
 equal to $G5$ and $A=7.3$ otherwise.

 A  first   sample has  been defined  with   Survey  stars by removing
 spectroscopic binaries (flag H59  set) and potential binaries in  the
 Hipparcos  solution  (flag H61  set   to S)  and keeping  stars  with
 relative errors on  parallaxes smaller than  10\% to avoid kinematics
 containing erroneous  or  additional motions.   The resulting  sample
 (11766   stars) has  been divided   into  19  sets  corresponding  to
 successive  ranges of  $B-V$ colour for  main sequence  stars, and  a
 twentieth set for giants in order to examine the kinematic variations
 between the subsets.

 We have used the proper  motions and parallaxes giving the tangential
 velocities to probe the   local kinematics.  Correction for  galactic
 rotation has been applied using   Feast \& Whitelock's  (\cite{fw97})
 values as done by Binney et al (\cite{bd97}).

 We also analysed a second sample in order to use the three components
 of velocity, adding  radial velocity  measurements to the  tangential
 velocities, and expecting  to  obtain a  stronger constraint on   the
 galactic structure.  However, it is known that radial velocities have
 been measured preferentially for high proper motion stars (see figure
 2   from Binney et al,  \cite{bd97})  and,  as a consequence, without
 careful  selection or analysis,   such  data may  introduce kinematic
 bias.  That  reason alone justifies  the necessity to analyse firstly
 the Hipparcos proper motions without radial velocities.

 In   order  to  analyze    the   full  three  kinematic   components,
 $v_r,~v_\theta,~v_z$, we  have used stars   from the McCormick survey
 (Vyssotsky, \cite{v63}).   This  sample   (895  stars)  is   based on
 spectroscopic surveys of  identification  of  M  dwarfs  and must  be
 without kinematic bias  (Weiss   \&  Upgren,  \cite{wu95}).    Radial
 velocities are obtained from the  literature; most data were found in
 the following  catalogues:  HIC (\cite{Hic}),  and more  recently  in
 Kamper  et al (\cite{kkt97}),  Reid et  al  (\cite{rhg95}), Upgren \&
 Harlow (\cite{uh96}).   The  Simbad data   base   has allowed  for  a
 systematic search  and cross-identifications.  Keeping only  data for
 which  velocity accuracy    is better than   $10\kms$   and rejecting
 identified spectroscopic binaries  and variables, we obtain  a sample
 of  309 stars.  These stars have  kinematics slightly hotter than the
 sample of giants.  They have been combined to  a sample of giants and
 red stars from HIP survey stars.

 \subsection{Preliminary analysis} 
 To    understand the main   behaviour   of the distribution  function
 $f(u,v,w)$ for the various stellar samples,  we refer to the analysis
 given  by Chereul et al (\cite{ccb97},  \cite{ccb98}),  G\'omez et al
 (\cite{gg97}),  Figueras     et  al   (\cite{fg97}),  Binney    et al
 (\cite{bd97}), Dehnen (\cite{d98}),  giving  both the axis   ratio of
 velocity distributions and descriptions of clumps  in phase space and
 irregularities.  We have  proceeded  to a Schwarzschild  distribution
 fitting  (more  exactly to  a  3D gaussian  fitting  with  one of the
 principal axes  fixed to z axis)  to the Hipparcos proper motions and
 ($l, b$)  coordinates  distribution\footnote{ A similar  analysis has
 been done by Dehnen \& Binney (\cite{db98})  who use a closed form to
 invert the Hipparcos tangential motions  into mean solar velocity and
 velocity dispersions.  The clear advantage    of their method is   to
 avoid  any prior on  the underlying  velocity distributions (Gaussian
 for instance).   While our classical Gaussian  fitting give {\it mean
 velocities}  of stellar groups close   to the values  they obtain, we
 obtain    different   results   concerning   the {\it   dispersions}.
 Performing numerical simulations we find that their method introduces
 a bias  on estimated dispersions.  The  bias increases with the ratio
 of the mean  velocity to the dispersion and  cannot be neglected  for
 the $\sigma_v$ dispersions.  }.

\begin{figure}[tbp]
   \begin{center}
 \centerline{\epsfig{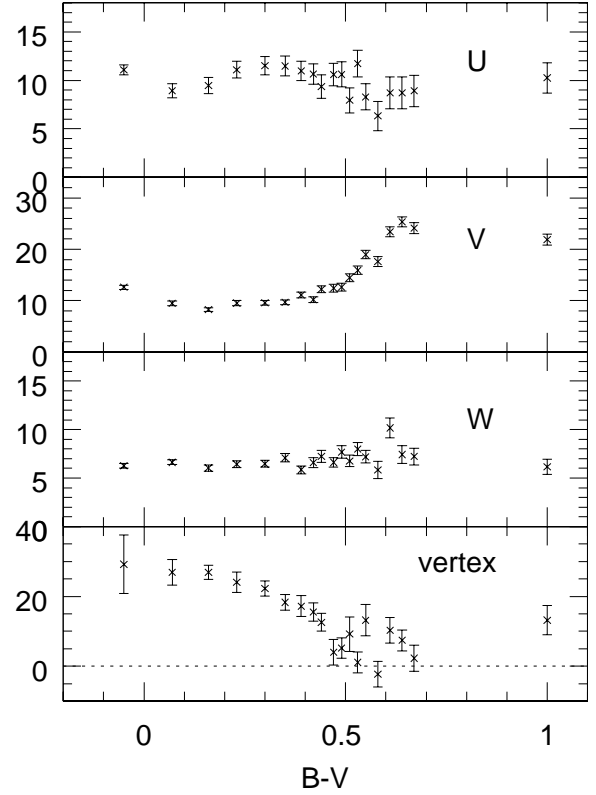}}
   \end{center} \caption{Mean velocity  ($\kms$) and  vertex  deviation
   dependences on stellar groups in colour bins.}
 \label{fig1}
 \end{figure}

\begin{figure}[tbp]
   \begin{center}
 \centerline{\epsfig{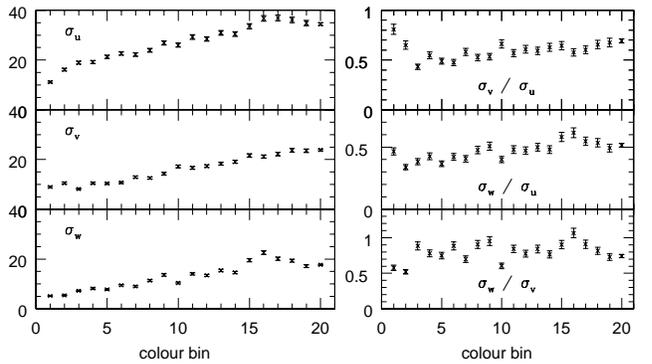}}
   \end{center}
   \caption{Velocity  dispersions and ratios. }
 \label{fig2}
 \end{figure}

\begin{figure}[tbp]
   \begin{center}
 \centerline{\epsfig{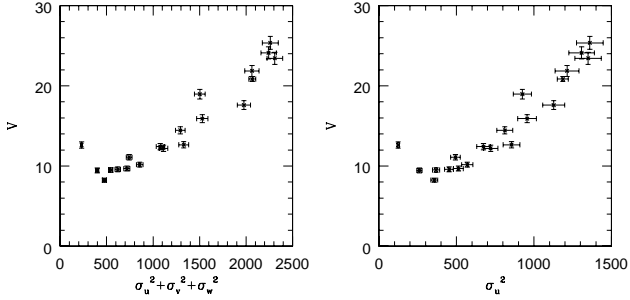}}
   \end{center}
   \caption{Tangential  velocity drift versus velocity variance.}
 \label{fig3}
 \end{figure}

 Table~\ref{tab1} and   Figs.~\ref{fig1}-\ref{fig2}-\ref{fig3}    give
 results for each  colour bin. U is  the velocity towards the Galactic
 center, V towards   the Galactic  rotation and  W  towards  the north
 Galactic  pole.   Fitting with two ellipsoids  does   not improve the
 determination,  the second  fitted ellipsoid   adjusting one  of  the
 moving group.  We   find the solar   velocity relative  to the  local
 standard  of   rest   (LSR)   to  be  $U_\odot=9.90\pm0.20\kms$   and
 $W_\odot=7.05\pm0.10\kms$.  The  $V_\odot$ component is usually taken
 as  the limit  $\sigma_u^2 \rightarrow   0$  in the  asymmetric drift
 relation  (Fig.~\ref{fig3}).  However  the slope is  not constant and
 the limit may range between  4 and $8 \kms$  depending heavily on the
 range  of  $\sigma_u^2$   variances  selected.  The  fact    that the
 asymmetric drift  relation is  not  linear cannot be related  to  the
 $\sigma_w/\sigma_u$  variation   that  is too  small.   The  simplest
 explanations  should be  that  the  density or  the   kinematic scale
 lengths change  gradually   with  populations  or that   the   bluest
 populations  are not kinematically relaxed.   Such  features are also
 marginally visible in Fig.~1 given by Mayor (\cite{m74}).

 The vertex deviation is maximum  for  blue stars ($27\degr$) and  for
 the reddest stars it fluctuates around $6\degr$.  Since red stars are
 a mixture of old and young stars with a large vertex deviation, it is
 compatible  with a null vertex  deviation for the old stars. Accurate
 ages  help  in answering  this  question and   show a null  deviation
 (G\'omez et al, \cite{gg97}).

 The velocity  ellipsoid axis ratio $\sigma_w/\sigma_u$ varies rapidly
 with colour bins, increasing from 0.45 for blue  stars to 0.55 to red
 ones, similar to the feature  obtained by G\'omez et al (\cite{gg97})
 by splitting stars by age intervals.

  \begin{table}[tbp]

    \caption{Velocity    dispersions ($\kms$)     and vertex  deviation
    (degree) for the twenty  stellar groups.  Colour  bins 1 to 18 have
    500 stars, the $19^{th}$ 477 and giants groups, the $20^{th}$, 2208
    stars.}

    \label{tab1}
     \begin{center}
      \begin{tabular}[tbd]{rcccccccc}
        \hline \\

 bin & $<B-V>$ & $\sigma_U $ & $ \sigma_V $  & $ \sigma_W $ & $vertex$ \\

        \hline \\
   1 &-0.05 & 11.1 &  9.0 &  5.2 & 29.2 \\
   2 & 0.07 & 16.1 & 10.5 &  5.5 & 26.9 \\
   3 & 0.16 & 18.9 &  8.2 &  7.2 & 26.9 \\
   4 & 0.23 & 19.2 & 10.5 &  8.2 & 24.1 \\
   5 & 0.30 & 21.3 & 10.4 &  7.8 & 22.3 \\
   6 & 0.35 & 22.6 & 10.7 &  9.5 & 18.3 \\
   7 & 0.39 & 22.2 & 12.9 &  9.0 & 17.2 \\
   8 & 0.42 & 23.9 & 12.5 & 11.4 & 15.5 \\
   9 & 0.44 & 26.8 & 14.3 & 13.6 & 12.6 \\
  10 & 0.47 & 26.0 & 17.3 & 10.4 & ~4.0 \\
  11 & 0.49 & 29.2 & 16.6 & 14.0 & ~5.2 \\
  12 & 0.51 & 28.5 & 17.3 & 13.4 & ~9.2 \\
  13 & 0.53 & 30.9 & 18.3 & 15.4 & ~1.1 \\
  14 & 0.55 & 30.4 & 19.1 & 14.6 & 13.2 \\
  15 & 0.58 & 33.6 & 21.6 & 19.5 &-~2.3 \\
  16 & 0.61 & 36.7 & 21.2 & 22.6 & 10.3 \\
  17 & 0.64 & 36.9 & 22.2 & 20.2 & ~7.4 \\
  18 & 0.67 & 36.2 & 23.7 & 19.3 & ~2.3 \\
  19 & 1.00 & 34.8 & 23.5 & 17.2 & 13.2 \\
  20 & giants & 34.4 & 23.8 & 17.7 & ~9.2 \\
       \hline \\
       \end{tabular}
       \end{center}
       \end{table}

 \subsection{Model fitting, parameter estimates and errors}

 The  Schwarzschild  distribution in this   section  and the dynamical
 models  in next    ones  represent  velocity  distribution  functions
 $f(v_r,v_\theta,v_z)$  for stellar  disc populations.   The predicted
 tangential velocities distribution $f(l,b; T_l,T_b)$ as a function of
 galactic coordinates  $l,b$ and tangential  velocities $T_l,T_b$  are
 easily deduced.  Stars with  extreme  velocities larger than 3  times
 the mean dispersions have    been rejected (11685  stars  remaining).
 This allows for the  removal of possible halo  stars or  objects that
 could not  be described by our   disc star modeling.   All models are
 fitted to data   by  adjusting the  model   parameters by  a  maximum
 likelihood.  This means that   the fitting procedure   gives unbiased
 estimates of the model  parameters (and is asymptotically convergent)
 {\it  if}  the observations   are drawn  from  a  parent distribution
 corresponding to the  assumed  model.  Errors  are obtained  from the
 second derivatives of the likelihood at the position of maximum.  The
 inverse  of this  matrix  is the  covariance   matrix of the  derived
 parameters giving errors and  correlations. The estimated errors have
 also been checked by repeating fits to random subsamples.

 Thus  errors  on  parameter estimates   are   dominated  by the  data
 sampling.  A much smaller source of errors comes from the accuracy of
 proper motions and distances measured by Hipparcos.  A crude estimate
 of  such  errors is obtained   by considering  the  median  error  on
 tangential  velocities  (from  proper  motions, parallaxes and  their
 respective errors  within our sample) and  we find $1.3 \kms$ for one
 velocity component.   With a sample of  11\,600  stars, the resulting
 error, for instance, on each  solar velocity component is only  $0.01
 \kms$.   On the other  hand, systematic effects, such as correlations
 between the Hipparcos solutions for close stars, exist while at large
 scale they must be null at the exception of  the error related to the
 link to  an inertial reference  system.  The deviation from inertial,
 non-rotating within  $\pm0.25\,{\rm  mas/yr}$ (Kovalevsky  et al,
 \cite{kl97}), has negligible effects on the estimate of moments.

 Systematic errors  or bias due  to  the  inadequacy  of the model  are
 naturally the most difficult  to evaluate.  Only  a more general model
 would help  to  quantify that  question.   For  instance the  non-zero
 vertex deviation is  not  considered by   Eq.~1  the axisymmetric model
 (a non-axisymmetric  model    with  three  integrals   could  have  been
 considered,  but in  the case  of   a St\"ackel potential,  the vertex
 deviation would have been the same for all stellar groups).  Certainly
 the vertex deviation  introduces a  bias  on the estimate of  velocity
 moments if the moments are estimated  assuming a null-deviation.  This
 will result in a bias on the other estimated parameters, scale length,
 slope of the rotation curve and is discussed further.

 \begin{sloppypar}
 A non-parametric approach  could seem  more attractive, avoiding  the
 bias of parametric models    and  allowing for a   possibly  unbiased
 determination  of the various  velocity   moments.  However, in  this
 paper,  we  want to avoid    a direct determination  of the  velocity
 moments   and then the  use of  relations  drawn from Jeans equations
 since we have shown that such equations lead to bias  (see \S 3.2 and
 Bienaym\'e \& S\'echaud   \cite{bs97}, see also  Kuijken \& Tremaine,
 \cite{kt91})  and that a more   rigorous  approach is acheived  using
 solutions of the Boltzmann equation.
 \end{sloppypar}
 \section{Kinematic model of galactic stellar discs}
 The   following analysis   is based   on a parametric    model for an
 axisymmetric thin   rotating stellar disk.   Such  model  was firstly
 introduced by     Shu (\cite{s69})  and   extended in   Bienaym\'e \&
 S\'echaud (\cite{bs97})  adding independent gradients for the density
 and the kinematics (see also Kuijken \& Dubinski, \cite{kd95}).  This
 model is a parametric  distribution function having a two-integral of
 motion  dependence, the  energy $E$ and   the angular momentum $L_z$.
 The fundamental  parameters are closely  linked to the radial density
 gradients of the disc stellar populations and also to their kinematic
 gradients and to  the shape of the circular  velocity curve.   It has
 been   applied  to the   analysis of  the   CNS3 (Gliese  \& Jahreiss
 \cite{gj91}) in  order    to  explain the observed    local  velocity
 distribution (Bienaym\'e \& S\'echaud, \cite{bs97}).

 The main  improvement of this  model  compared to Shu's  model is its
 ability to reproduce  the observed marginal velocity  distribution of
 stars.  This  is     due to  the  introduction   of   three realistic
 independent  parameters,   the  density,   kinematic dispersion   and
 rotation curve gradients  (see also  previous investigations of   the
 local kinematics with dynamically  coherent distribution functions by
 Kuijken \& Tremaine, \cite{kt91},  Evans \& Collett,  \cite{ec93} and
 Cuddeford \& Binney \cite{cb94}).

 Based on  an application  of  the  Jeans equation,  Fux  \&  Martinet
 (\cite{fm94}) deduced from the observed moments (asymmetric drift and
 velocity dispersions) that the density scale length  is close to $2.5
 -3.0  \kpc$  with  assumptions  on the  closure  of  the hierarchy of
 velocity moment equations   (Cuddeford \& Amendt, \cite{ca92}).    We
 must however notice that such  analysis using first and second  order
 moments of the distribution  neglects available informations like the
 velocity distribution kurtosis and skewness.

 The galactic potential can be fitted by a  St\"ackel potential in the
 solar neighbourhood.  There the  effective potential can  be expanded
 in powers of (r-$R_0$) and z.  How well can this expansion be matched
 by a similar one  for a separable potential?   For instance,  a first
 order development  gives a $r$ and $z$   separable potential which is
 then used for  the estimate of the force  perpendicular to the  plane
 and for the mass  density determination in  the plane.  Van  de Hulst
 (\cite{vh62}) and Kent \& de Zeeuw (\cite{kz91}) solved the expansion
 problem with St\"ackel potentials.   The expansion includes the third
 order  term  $rz^2$ related to    the potential flattening.   If  the
 galactic potential were  of St\"ackel  form we  would have an   exact
 measure of its flattening (through the parameter  $z_0$, the focus of
 the  ellipsoidal  coordinate system defined   in the  Appendix).  The
 galactic potential is certainly not exactly  of a St\"ackel form, but
 probably not so far from it  as long as  like here, we consider stars
 within a  small range of excursion  from the solar position ($3 \kpc$
 radially and $1 \kpc$  vertically).   A more general  analysis (using
 orbit computations for  instance)   could  establish accurately   the
 coupling between the vertical and radial  motions and its link to any
 potential.    The   advantage   of   St\"ackel   potentials  is their
 tractability and they are certainly sufficient  for a first analysis.
 In this paper  it allows us  to establish formally the importance  of
 the potential  flattening   on the  stellar    motions in the   solar
 neighbourhood.

 \subsection{3D axisymmetric stellar disc kinematic model}

 The model is a 3D extension of the 2D distribution function described
 in Bienaym\'e  \& S\'echaud (\cite{bs97}).  This  model is defined in
 3D axisymmetric St\"ackel  potentials where three integrals of motion
 are  known  ($E$, $L_z$,  $I_3$).  Being  a  function  of these three
 integrals,   $f(E,L_z,I_3)$   is   a   stationary   solution  of  the
 collisionless  Boltzmann   equation.   It   is also   close    to the
 Schwarzschild  distribution  for small  velocity  dispersions and its
 associated density is nearly exponential as  are the kinematic radial
 distributions.

 The expression  of the distribution  function is given by (see also the
 Appendix for more details):
 \begin{equation}\begin{split}
 f(E,L_z, I_3) =
 \frac{2\Omega(R_c)}{ 2\pi\kappa(R_c)}
 \frac{\Sigma(L_z)}{
            \sigma_r^2(L_z) }
 \exp \left[
 -\frac{E-E_{circ}}{\sigma_r^2}
 \right] \\
 \frac{ 1}{ \sqrt{2\pi} } \frac{1}{\sigma_z(L_z) }
 exp \left[
 -(\frac{R_c(L_z)^2 }{ z_o^2 } +1 )^{-1}
 (\frac{1}{\sigma_z^2}-\frac{1}{\sigma_r^2})I_3
 \right]
 \end{split}
 \end{equation}
 and is  null when  $L_z<0$, where  $R_c=R_c(L_z)$   is the  radius of
 circular orbit stars  with   angular momentum  $L_z$ and having   the
 energy $E_{circ}(R_c)$.   $\Omega$ is the angular  velocity, $\kappa$
 the epicyclic frequency and
 \begin{equation}
 \Sigma(L_z) = \Sigma_0 \exp[-R_c(L_z)/R_\Sigma]
 \end{equation}
 \begin{equation}
 \sigma_r (L_z) = \sigma_{0,r} \exp[-R_c(L_z)/R_{\sigma_r}].
 \end{equation}
 \begin{equation}
 \sigma_z (L_z) = \sigma_{0,z} \exp[-R_c(L_z)/R_{\sigma_z}]~,
 \end{equation}
 in the following, we put $R_{\sigma_z}=R_{\sigma_r}$ and we define the
 radial dependence of the potential  in the  galactic plane assuming  a
 power law rotation curve $v_c(r)\sim r^{\alpha}$.

 \subsection{some properties}

\begin{figure}[tbp]
   \begin{center}
 \centerline{\epsfig{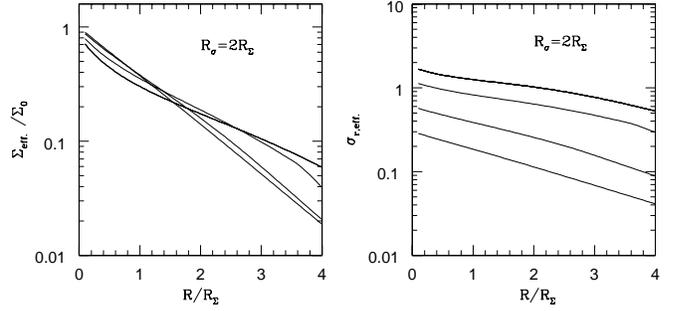}}
   \end{center}
   \caption{Effective  density and  radial velocity dispersion computed
   from the corresponding  moments of  the distribution function  Eq.~1
   (models computed  with   $z_o=\infty$ and   a flat   rotation  curve
   $v_c(r)=1$, $\sigma_{r,eff}$ are   plotted in the  same  units).  Four
   radial density  distributions with  $R_{\sigma}=2R_{\Sigma}$     and
   $\sigma_0=.3-.6-1.2-1.8$ are plotted.      Solar   position     is  at
   $R_0/R_{\Sigma}=2.5-3.5$  and $\sigma_r( R_0/R_{\Sigma})=0.3$ for the
   thick disc.  Above a   critical $\sigma_0$ value, the density  scale
   length increases and is no longer close to $R_\Sigma$.}
 \label{fig4}
 \end{figure}

 For  a convenient range  of  parameters ($R_\Sigma$, $R_{\sigma_r}$),
 such a distribution   has effectively nearly  exponential density and
 kinematic decreases with respective scale lengths close to $R_\Sigma$
 and $R_{\sigma_r}$  (Fig.~\ref{fig4}).   Models with  extremely large
 velocity  dispersions cannot  have small   scale lenghts (i.e.  large
 density  gradients).  In such    cases, the effective  density  scale
 lengths are different from   the input parameter $R_\Sigma$  and such
 models no  longer have an exponential  decrease over a large range of
 radius (Fig.~\ref{fig4}).  Analysing the  Hipparcos data, we will not
 be concerned by this  problem since the velocity dispersions involved
 in the data are small.

\begin{figure}[tbp]
   \begin{center}
   \centerline{\epsfig{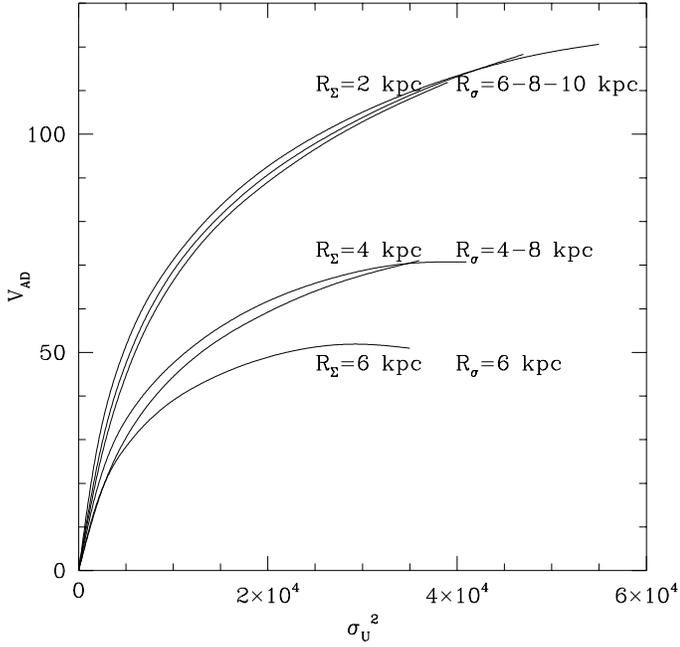}}
   \end{center}
   \caption{Asymmetric   drift  relations.   Tangential  velocity drift
   versus radial velocity variance for  models with various density and
   kinematic gradients ($R_0=8.5 \kpc$, $V_0=220\kms$). Models with the
   largest  density scale length  present   a saturation effect on  the
   amplitude of the tangential velocity drift.  }
 \label{fig5}
 \end{figure}

 The asymmetric   drift (AD) relates  the  velocity dispersion  of any
 stellar group to its tangential drift.  A quite general form is given
 by Binney  \& Tremaine,  \cite{bt87} (BT) (Eq.   4.34), valid  in the
 galactic plane  at $z=0$, relating linearly  $V_{AD}$ to the variance
 $\sigma_r^2$  (for  constant dispersion   ratios and kinematic  scale
 length that is  independent of stellar  types).  That equation can be
 considered  as a first   order  development with a  limited  range of
 validity.   We  have  computed  the   asymmetric  drift relation  for
 stationary solutions given by our  model and the results are  plotted
 (Fig.~\ref{fig5}) for different  sets of parameters.  The AD relation
 is found to  be nearly linear in the  range of drift  from $0$ to $30
 \kms$.  This interval is   apparently the range  of validity   of the
 classical equation.

 Hipparcos  data as well  as other data   used to study the asymmetric
 drift fall in  this  range of  velocity  drift.  However, as  seen in
 \S2.1, the asymmetric drift relation is not strictly linear.  Finally
 for  stellar populations having larger drifts,  like  the thick disc,
 the use of a linear relation is certainly wrong.

\begin{figure}[tbp]
   \begin{center}
 \centerline{\epsfig{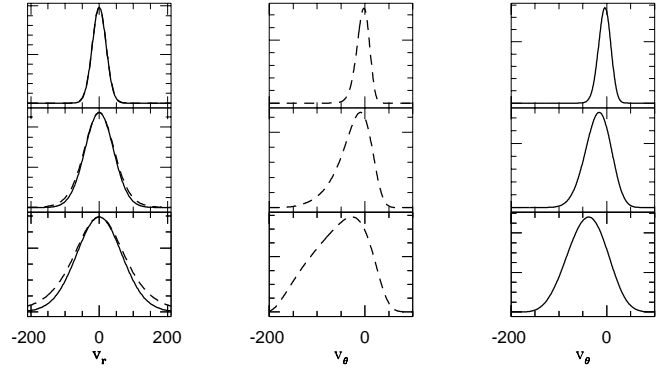}}
   \end{center} \caption{Marginal velocity  distributions
   at  solar position, for  two models with short  (full lines) and long
   (dashed lines) density scale lengths.  Top to bottom with increasing
   velocity dispersions.   Left: radial velocity  distribution for both
   models, middle:  tangential  velocity  distribution for  model  with
   $R_{\Sigma}=2.5\kpc,~R_{\sigma}=7.5\kpc$, right:  for  model   with
   $R_{\Sigma}=1.8\kpc~R_{\sigma}=15\kpc$.}
 \label{fig6}
 \end{figure}

 The asymmetric drift  is  neither entirely  described by the   moment
 equation 4.34  of  Binney \&  Tremaine (\cite{bt87}) nor  by the more
 exact relations such these plotted on  Fig.~\ref{fig5} since, in some
 range  of drift, different modeling   may have similar AD  relations.
 Higher  moments or marginal velocity  distributions  help to identify
 and    discriminate   between   models.     This  is   recognized  in
 Fig.~\ref{fig6} where  various   signatures,   like drift  but   also
 kurtosis and  skewness     of the    $v_\theta$ distributions,    are
 identifiable.  Differences due to a short or  a long scale length are
 visible.  In the  case of the shortest  scale length $1.8\kpc$,  with
 $\sigma_r$ increasing, $f(v_\theta)$ stays  symmetric and the maximum
 shifts  regularly.   For     larger   scale length  $2.5\kpc$,    the
 $f(v_\theta)$  distribution  becomes   highly  asymmetric.  At higher
 scale  length $4.5\kpc$, the maximum  of the $v_\theta$ distributions
 is nearly not shifted.

 \subsection{$f(E,L_z,I_3)$ at $z=0$}

 We give, below, the simpler expressions  of the distribution function
 in the galactic plane ($z=0$)  (see Appendix) corresponding closer to
 the situation of the Hipparcos sample we are analysing. We have:
 \begin{eqnarray}
 f(E,L_z,I_3)=
 g(r;\dot{r},\dot{\theta})
 f_{\perp}(r,z;\dot{\theta},\dot{z})
 \end{eqnarray}
 with
 \begin{eqnarray}
 f_{\perp} =
 \exp \left\{
 -\frac{\dot{z}^2}{2}
 \left[ \frac{1}{\sigma_r^2}+(\frac{1}{\sigma_z^2}-\frac{1}{\sigma_r^2})
 \left( \frac{ r^2+z_o^2 }{R_c(L_z)^2 +z_o^2} \right) \right]\right\}
 \end{eqnarray}
 where  $z_o$ (defined in Appendix  as  the focus  of the  ellipsoidal
 coordinate system) is related to the shape of a St\"ackel potential.

 For a flat rotation curve with $v_c(r)=v_o$
 \begin{eqnarray}
 f_{\perp} =
 \exp \left\{
 -\frac{\dot{z}^2}{ 2}
 \left[ \frac{1}{\sigma_r^2}+(\frac{1}{\sigma_z^2}-\frac{1}{\sigma_r^2})
 \left( \frac{ r^2 v_o^2+z_o^2  v_o^2}{ r^2 v_\theta^2+z_o^2  v_o^2}
\right) \right]\right\}.
 \end{eqnarray}

 This $v_\theta$, $v_z$ coupling is established only in  the case of a
 St\"ackel  potential   and  of  a   quasi-exponential  stellar  disc.
 Different couplings would  have  been obtained  if  we had considered
 another density distribution.

 Towards $z_o=\infty$, the potential is cylindrical and
 \begin{equation}
 f_{\perp} =
 \exp \left\{ -\frac{\dot{z}^2}{ 2\sigma_z^2} \right\}.
 \end{equation}
 Towards $z_o=0$, the potential is spherical and
 \begin{equation}
 f_{\perp} =
 \exp \left\{
 -\frac{\dot{z}^2}{ 2\sigma_z^2}
 \left[\frac{\sigma_z^2}{\sigma_r^2}+(1-\frac{\sigma_z^2}{\sigma_r^2})
 \frac{v_o^2}{ v_\theta^2} \right] \right\}.
 \end{equation}

 Dependence on the  $z_o$ parameter is  easily identifiable and can be
 isolated from the other parameter dependencies.  This is discussed in
 \S5 and  in the   Appendix.    This  dependence links   the  velocity
 distribution to the potential, bringing  a constraint on the galactic
 mass distribution.

 \section{Model versus data: solar motion and galactic scale lengths}

  \begin{table}[tbp]

    \caption{Solar  motion   relative     to  the  LSR   and   galactic
    characteristics   deduced  from   the   local  stellar   kinematics
    (1-$\sigma$ errors).   Best  estimates of parameters  are shown
    for two ranges of data (2 to 20 and 3 to 20  bins) and with models
    with one kinematic   scale   length  $R_{\sigma_r}$  or  with   two
    differents scales for blue and red stars.  Typical errors are given
    in the last column but do not  include the correlations. Except for
    $u_\odot$  and $w_\odot$, they are  raised by a factor $\sim 1.5-2$
    when correlations are considered.}

    \label{tab2}
     \begin{center}
      \begin{tabular}[tbd]{ccccc}
 ~& bins 2-20 & bins 3-20 & bins 2-20 & bins 3-20 \\
 ~& one $R_{\sigma_r}$ & one $R_{\sigma_r}$ & two $R_{\sigma_r}$ & two
$R_{\sigma_r}$ \\
        \hline \\
 $u_\odot$  &9.5 & 9.5 & 9.6 & $9.7\pm.3$ \\
 $v_\odot$  &3.8 & 3.5 & 5.2 & $5.2\pm.5$ \\
 $w_\odot$  &6.7 & 6.7 & 6.7 & $6.7\pm.2$ \\
 $R_\Sigma $&1.5 & 1.5 & 1.7 & $1.8\pm.1$ \\
 $R_{\sigma_r}$ &10.0 & 10.0 & 9.3 \& 19 & $9.7\pm.4~\&~17\pm2$\\
 $\alpha$ & -.22 & -.24 & $-.22$ & $-.24\pm.02$ \\
 $z_o $ & 19 & $17$ & $23$ & $24\pm10$\\
        \hline \\
       \end{tabular}
       \end{center}
       \end{table}

 Model to data comparison is achieved  by splitting the stellar sample
 by colour intervals covering various ranges of kinematics.  The ratio
 between the widths     of the marginal   distribution  in   $v_r$  to
 $v_\theta$ is mainly linked to the slope of the rotation curve, while
 the shape   of   the $v_\theta$  marginal distribution    is strongly
 correlated  to  the scale lengths.    The exact combination  of these
 correlations is untangled by  fitting the model  distribution (Eq.~1)
 to the observed velocity distributions.

 \subsection{Results of model fitting to Hipparcos tangential velocities}

 Results are given in Tab.~\ref{tab2}.   For each colour bin we adjust
 one distribution function (Eq.~1) with the common free parameters for
 all colour bins  being the solar motion, $R_\Sigma$,  $R_{\sigma_r}$,
 $\alpha$   (logarithmic  slope of  the   rotation  curve) and  $z_o$.
 Parameters not given in  Tab.~\ref{tab2} are the radial  and vertical
 velocity dispersions associated to each colour bin.

 We find  that the   determination  of the tangential  solar  velocity
 $v_\odot$ is sensitive to the first colour  bin (youngest stars) that
 we have rejected.  The density scale length is also correlated to the
 solar  velocity $v_\odot$. All the   results   favour a short   scale
 length.

 Since the AD  relation  is not linear  (Fig.~\ref{fig3}), it   is very
 natural to consider that the $R_\Sigma$, $R_{\sigma_r}$ parameters are
 not the same for all stellar populations.  The change of the ellipsoid
 axis ratio cannot explain the bending of the AD relation.  In external
 galaxies the  change with colour of  the density  scale length is only
 about 20\%   (de Jong, \cite{dj96}).  However  the  youngest stars may
 have  a smaller density scale  length close to the  ISM one, though it
 also  depends  on the  SFR   radial dependence.  More  certain,  the
 kinematic scale length  of youngest populations  must be  close to the
 ISM  kinematics with a nearly  null  gradient.  The different rate  of
 dynamical heating with  radius  should change  the kinematics  to  the
 observed one for old populations.  For  these reasons, we have tried a
 modeling with different  kinematic scale lengths  for the bluest stars
 (bins 3-10) and for the reddest stars (bins 11-20).  The likelihood is
 improved; we find out $v_\odot= 5.2\kms$  and the density scale length
 $1.8\kpc$.  We   find for the   bluest stars a nearly   null kinematic
 gradient, in accordance  to  our expectation,  and for reddest   stars
 $R_{\sigma_r}$ is $9.7\kpc$ (or $R_{\sigma_r^2}=4.9\kpc$) close to the
 Lewis \& Freeman 's value (\cite{lf89}) $R_{\sigma_r^2}=4.4\kpc$ based
 on 600 giants towards the centre and anticentre directions.

 The  comparison of the maximum  likelihood  (ML) of respective models
 allows us to determine their relative goodness of fit: the likelihood
 ratio test    giving  a measure of   the   reliability of adding  new
 parameters.    The  likelihood-ratio   statistics   (Nemec   \& Nemec
 \cite{nn91}, Kendall    \&  Stuart    \cite{ks67})  is   $\lambda   =
 \frac{{\cal{L}}_1}{{\cal{L}}_2}$    where   ${\cal{L}}_1$ is      the
 likelihood for  models  based  on  $K_1$ components.  Models   with a
 larger number  of   components  $K_2$ may  give   a better  fit  with
 $\lambda$  smaller  than  one.   The  statistical   significance   of
 $\lambda$    is obtained by   comparing $-2\log\lambda$  to the upper
 percentile of  a $\chi^2$ distribution with the  number of degrees of
 freedom equal to  the difference in  the number of parameters between
 the two models.  (However, these results only  hold if the asymptotic
 normality and efficiency of  the ML estimator are  satisfied, Kendall
 \& Stuart, \cite{ks67}).
  \begin{table*}[tbp]

    \caption{Logarithm of likelihoods of models. Likelihood ratio tests
    the improvement of the goodness of fits.}

    \label{tab-rep1}
     \begin{center}
      \begin{tabular}[tbd]{ccccc}
 models & ln (${\cal L}_{bins~3-20}$ )& Nb parameters & ln (${\cal
L}_{bins~2-20}$ )& Nb parameters \\
        \hline \\
 Gaussian fit                        & -93993.3 & 126 & -97753.3 & 133\\
 Eq.~1: two kinematic scale lengths & -94090.6 & 45  & -97872.4 &  47\\
 Eq.~1: one kinematic scale length  & -94093.2 & 44  & -97887.9 &  46\\
 gaussian fit with null vertex       & -94215.2 & 108 & -97992.8 & 114\\
        \hline \\
       \end{tabular}
       \end{center}
       \end{table*}
 Likelihoods of the discussed models are summarized in Tab.~\ref{tab-rep1}.
Summarizing:

 1)  We find that Gaussian  fits  are significantly  improved when  the
 vertex is adjusted ($\lambda \sim 220$ for $\sim 20$ more parameters).

 2)  Comparing the  Gaussian  fit with  a  non null  vertex   to Eq.~1
 modelings, we see that the   Gaussian gives a significantly  improved
 fit (and considering they have more free adjusted parameters).

 3)  More interesting is that when  Eq.~1 modelings are compared to the
 more  similar Gaussian fits with  null vertex, the  likelihood is very
 significantly improved considering  that the number of free parameters
 is lowered by about 65.

 4)  Finally comparing the two Eq.~1  models with  one or two kinematic
 scale lengths, the change of likelihoods is significant (as we add one
 supplementary parameter, the fit is improved) ($-2\log\lambda$=5.2 and
 30 for 18 and  19 bins, corresponding  respectively to an  improvement
 with a probability of 0.98 and $1.-5*10^{-8}$).

 Errors given  in the last column of  Tab.~\ref{tab2} are deduced from
 the diagonal of the  covariance matrix.  Existing correlations  raise
 the errors.  A good  indicator is the global  correlation coefficient
 (Eadie et al, \cite{ed71}) a quantity that is  a measure of the total
 amount of correlation between a  variable and all the other variables
 (this coefficient being one  if the variable  is a linear combination
 of the  other variables).  Most  correlations and global correlations
 are small except for the following variables: 1)  a correlation and a
 global correlation about  0.8  between $v_\odot$ and  $R_\Sigma$;  2)
 $R_{\sigma_r}$ has  correlations of  0.3 and  0.5 with  $v_\odot$ and
 $R_\Sigma   $ and  a  global correlation   of  0.8;  3) $\alpha$  has
 correlations of 0.2  and 0.3 with $v_\odot$  and $R_{\sigma_r}$ and a
 global correlation of    0.5.    This means that   errors    given in
 Tab.~\ref{tab2} must be multiplied  by $\sim 1.5-2$ for the variables
 $v_\odot, R_{\sigma_r}, R_\Sigma $ and $\alpha$.

 The estimated density scale length   is directly proportional to  the
 adopted solar  galactic radius $R_0$ (here  taken as $8.5 \kpc$).  It
 also depends on the adopted $V_0$ in a  less direct way. We find that
 a   decrease  of  $V_0$  by   10  percent increases  the  estimate of
 $R_\Sigma$ by about 10 percent while the slope  of the rotation curve
 is unchanged,  $\alpha=-0.23$.   $R_0$ and $V_0$  are  not accurately
 known but they are  linked by  the  determination of  the $R_0$/$V_0$
 ratio (from the Oort's constants  difference $A-B$), it results  that
 $R_{\Sigma}$ remains  nearly  unchanged  by   the  $R_0$  and   $V_0$
 uncertainties.

 It is   worth noting that   we  find a  decreasing   rotation curve in
 agreement with the new determination of the Oort's constants (Feast \&
 Whitelock,     \cite{fw97},   or    Olling   \&   Merrifield,    1998)
 ($A+B=(dV/dR)_0=-2.4\kms\kpc^{-1}$).     Considering   the constraints
 given by  the  rotation curve  (Rohlfs  et  al, 1988),  a   decreasing
 rotation curve is obtained only  with small  $R_0$ and $V_0$  (smaller
 than the IAU  recommended values).  Olling \& Merrifield (\cite{om98})
 find $R_0=7.1\kpc$ and $V_0=184\kms$.  The corresponding $\alpha$, the
 logarithmic slope   of the  rotation curve, is    then $-0.10$ not  in
 excellent agreement with our local determinations.

 This difference could be partly explained by the most evident bias of
 our model, i.e. the assumption of  a null vertex deviation that leads
 to biased  estimates of   the  moments of  the ellipsoidal   velocity
 distributions.  This bias can  be evaluated by comparing the velocity
 moments  determined with a  non-null   vertex deviation (Tab.~1)  and
 those determined  assuming a null deviation.   The largest axis ratio
 estimates are changed by 10 to 5 percent for the six first bins and 0
 to 3  percent for  the other bins.   We quantify  from Eq.~  4.52 and
 Eq.~4.35 (BT) the  possible  order of magnitude   of the bias  on the
 slope  of the  rotation.    $\alpha$ could  be  increased by  +0.1 to
 $-0.14$, while the estimate of scale lengths would be increased by 10
 percent.  Then  the  estimated  density scale  length  would be  $2.1
 \kpc$.  Our bias estimation cannot be applied  so simply, and we will
 consider that it  gives at least a more  realistic estimate of errors
 bars, much larger than the Poissonian errors.

 More  generally we remark that it  is certainly not clear how correct
 the use of equations like Eqs~4.35 or  4.52 (BT) obtained by assuming
 an axisymmetric galaxy is when  they are applied to observed velocity
 distributions that  do not satisfy the   symmetry conditions and when
 the  observed moments are  not computed  along the coordinate system.
 Limitation and  bias in the application  of our modeling  (Eq.~1) are
 probably also present when using Jeans and the related equations.

 These results are close  to previous kinematic determinations of  the
 scale lengths by Mayor  (\cite{m74}) ($R_\Sigma=2.2\kpc$) or Oblak \&
 Mayor (\cite{om87}) ($R_{\sigma_r}=10\kpc$ or $R_{\sigma^2_r}=5\kpc$)
 and  Fux \& Martinet (\cite{fm94})  ($R_\Sigma=2.5\kpc$) based on the
 local stellar kinematics.

 \begin{sloppypar}
 The   short density scale   length  we  have   determined is also  in
 agreement  with the following  determination from star  counts at low
 galactic latitudes: with IR star counts  by Porcel et al (\cite{p98})
 ($2.1\kpc$),    Ruphy et  al (\cite{r96})    ($2.3\kpc$), Kent et  al
 (\cite{k91}) ($3.0\kpc$)  and in visible  by Robin et al (\cite{r92})
 ($2.5\kpc$).  Higher latitude star  counts (in visible) like those by
 Siegel et al  (\cite{smr97}) may give  a longer  scale length.  Could
 these results be  biased by thick disc  stars?  We  just mention here
 that the larger thick disc scale length ($3\kpc$) measured by Ojha et
 al (\cite{ob96}) using star counts  and proper motions that allow one
 to separate accurately the stellar populations.
 \end{sloppypar}

 Accuracy on $z_o$ is extremely low since it depends on $v_\theta,~v_z$
 coupling   that is  not  well defined   by  proper motions  alone.  At
 $2\sigma$ all  $z_0$ values are possible.

  \begin{table}[tbp]

    \caption{Best fit models to the 3D stellar velocity distributions.}

    \label{tab3}
     \begin{center}
      \begin{tabular}[tbd]{rccc}
       \hline \\
 $R_\Sigma$ & $z_o$ & $v_\odot$ & log of \\
 $\kpc$     & $\kpc$ & $\kms$   & likelihood \\
       \hline \\
 4.0        & $7.9\pm2.0$ & $12.4$ &-34.75\\
 2.5        & $6.9\pm1.4$ & $ 9.5$ &-31.22\\
 2.0        & $6.1\pm1.4$ & $ 7.8$ &-30.02\\
 1.7        & $5.7\pm1.4$ & $ 6.1$ &-29.64\\
 1.5        & $5.4\pm1.4$ & $ 5.0$ &-39.80\\
 1.3        & $5.0\pm1.4$ & $ 3.4$ &-30.49\\
 1.1        & $4.9\pm1.9$ & $ 1.3$ &-31.84\\
       \hline \\
       \end{tabular}
       \end{center}
       \end{table}

 \subsection{3D velocity data}

 3D velocity data have been used  in order to improve the determination
 of $z_o$ because this  parameter depends on  the coupling of  velocity
 components.

 For that purpose we first tried to determine $z_o$ with a subset from
 the McC  survey  (309 stars)  but  the accuracy was  not sufficiently
 improved.  We   augmented  this sample  with  HIP survey   stars with
 existing radial velocities.  While McC  sample is free from kinematic
 bias, this is not the case  for Hipparcos stars with published radial
 velocities.  However the completeness (and absence of kinematic bias)
 is   achieved  for high velocity  stars  (Binney  et al \cite{bd97}).
 Since $z_o$ is exclusively constrained by  the highest velocity stars
 and is unconstrained by stars with velocities smaller than $20 \kms$,
 we expect a small kinematic bias.

 \begin{sloppypar}
 Combining in a single sample,  McC stars (309),  giants (1239 from bin
 20) and reddest dwarfs  (833 stars, $B-V > .525$) with radial velocities
 we find  that the optimal fittings  are achieved with a four-component
 model  (i.e.:  adding four    elementary distributions  (Eq.~1)  with
 different weights  $\Sigma_0$    and radial  and   vertical   velocity
 dispersions     $\sigma_{r,0}$,     $\sigma_{z,0}$).    Exploring  the
 maximum    likelihood space, with the parameters being    free or
 setting   the  various     parameters   with   values  obtained     in
 Tab.~\ref{tab2}, we arrive  at  the following conclusions:  with these
 data $R_\Sigma$, $z_o$ and $v_\odot$   are correlated and may vary  in
 the range of models given in Tab.~\ref{tab3}.  The accuracy and errors
 are  determined from  the diagonal    of  the covariance matrix   that
 reflects the errors  due to  the  data sampling.   On the other  hand,
 varying $\alpha$ in the range -0.3 to  -0.1 (with $V_\odot$ fixed) has
 nearly no effect  on  the other  adjusted  parameters.  In  fact  this
 sample alone (hot   kinematic)  is suited to constrain   the parameter
 $z_o$  but not  accurately    the  others.   Combining  results   from
 Tab.~\ref{tab3} to these obtained using HIP tangential velocities from
 survey stars  (Tab.~\ref{tab2}),  where the solar  velocity  $v_\odot$
 ranges  between  4 and $6\kms$,  and  $R_\Sigma$ about $1.8  \kpc$, we
 deduce that $z_o$ is $5.7\pm 1.4\kpc$.
 \end{sloppypar}

 The  $z_0$ dependence on $R_0$ is  such that $z_0/R_0$ does not depend
 on $R_0$.  As a consequence the parameter  $\Lambda$ (defined in \S 5)
 does not depend on $R_0$. Concerning the $V_0$  parameter, a change of
 10 percent produces a negligible change of $\sim1$ percent on $z_0$.

 The vertical $R_{\sigma_w}$ and radial $R_{\sigma_r}$ kinematic scale
 lengths have been supposed equal because the data  do not allow us to
 determine separately the two  scale lengths.  They have certainly the
 same (large) order of magnitude.  (The  kinematic radial scale length
 $R_{\sigma_r}$ is about 9 kpc for giants, Neese \& Yoss, \cite{ny88},
 Lewis et Freeman  \cite{lf89}, a value  large compared to the density
 scale length.   For  younger populations  the  kinematic scale length
 should be even larger).

\begin{figure}[tbp]
   \begin{center}
 \centerline{\epsfig{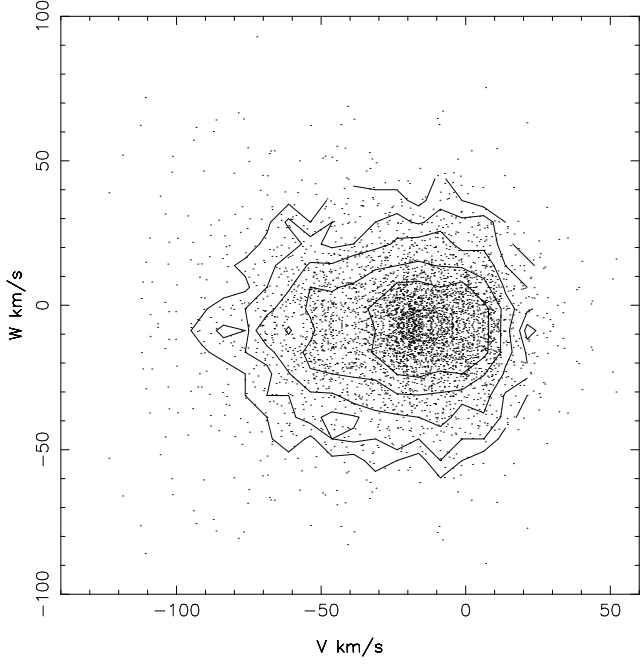}}
 \end{center}
   \caption{Symmetrized (axis $w=-7\kms$) observed $f(v,w)$ marginal velocity
   distribution.
 The  maximum   density corresponds  roughly to   Pleiades SuperCluster
 position (around v=$-22\kms$, w=$-5\kms$) and the innermost contour is
 certainly affected  by   non-stationary  features that  cannot  be
 modeled by  the stationary solution given by  Eq.~1:  it is the region
 where selection effects  are present but  it is also the region  where
 data do not constrain the parameter $z_0$.
 We notice the egg-shape   of  isocontours pointing towards   negative
 V-velocity.
 }
 \label{fig7}
 \end{figure}

\begin{figure}[tbp]
   \begin{center}
 \centerline{\epsfig{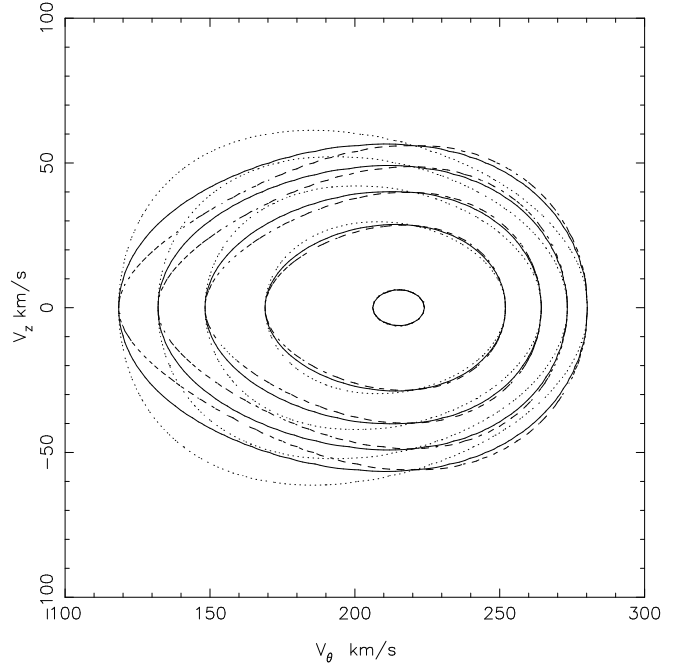}}
   \end{center}   \caption{Isocontours   of  the     $f(v_\theta,v_z)$
   distribution  for three models:  best  fit model (continuous lines),
   and best  fit model changing   $z_o$ to 0  (dashed) and  to $\infty$
   (dotted).
   Isocontours are egg-shape,  best fit and $z_0$=0 solutions pointing
   towards  negative    $V_{\theta}$-velocity   like  data  isocontours
   (Fig.~\ref{fig7}).   $z_0=\infty$  solution (dotted) has isocontours
   pointing towards positive $V_{\theta}$-velocity.}
 \label{fig8}
 \end{figure}

 Now  considering the  distribution function  (Eq.~1),  a  decrease of
 $R_{\sigma_w}$ (at    constant  $R_{\sigma}$)   has   the   effect of
 increasing the vertical dispersion for  velocities smaller than $V_0$
 and       decreasing  for    velocities       larger     than   $V_0$
 (Figs.~\ref{fig7}-\ref{fig8}).    The apparent    change   on     the
 isocontours would  mimic a change of the  $z_o$ parameter.  We notice
 it also changes the  velocity distribution $f(v_{\theta}, v_z=0)$ and
 then the two effects (changes on $R_{\sigma_w}$  and $z_o$ ) could be
 distinguished.   Finally, modifying  independently both  $R_{\sigma}$
 and  $R_{\sigma_w}$ in the range 6-15~  kpc,  gives a range of values
 for fitted $z_o$ from 3.8 to 6~kpc that is nearly  about the range of
 the errors given by the maximum likelihood.

 \section{Local constraints on the galactic potential}

 In building the 3D kinematic  model, we   defined the parameter $z_o$
 that is a measure of the radial  bending of the  potential. It is also
 tightly related to  $\delta$   the  vertical  tilt of the     velocity
 ellipsoid given by
 \begin{equation}\begin{split}
 \tan 2 \delta = \frac{ 2~\sigma_{r,z}^2 }{ \sigma_{r,r}^2-\sigma_{z,z}^2 }.
 \end{split}\end{equation}
 Hori \& Liu (\cite{hl63}) shown that in a St\"ackel potential the tilt
 $\delta$ depends only on the  positions:
 \begin{equation}\begin{split}
 \tan 2 \delta  = \frac{2~r~z }{ r^2 -z^2 + z_o^2 }.
 \end{split}\end{equation}
 The  change of  the tilt  close  to  the galactic plane  (where $z=0$,
 $\delta=0$ and first order vertical  derivatives of $\sigma_{r,r}$ and
 $\sigma_{z,z}$ are null) is
 \begin{equation}\begin{split}
 \partial_z  (\tan    2   \delta   )  =    \partial_z   \sigma_{r,z}^2~
 \frac{2}{\sigma_{r,r}^2-\sigma_{z,z}^2 } = \frac{\Lambda(r)}{ r}
 \end{split}\end{equation}
 and for a St\"ackel potential (at $z=0$)
 \begin{equation}\begin{split}
  \Lambda(r)= \frac{r^2 }{ r^2 + z_o^2 }.
 \end{split}\end{equation}

 A general and approximate expression valid at $z=0$ (exact in the case
 of St\"ackel potentials) is given by Amendt \& Cuddeford (\cite{ac91})
 (see also    Cuddeford  \&  Amendt, \cite{ca91},    \cite{ca92}) where
 $\Lambda(r)$ depends on potential derivatives:
 \begin{equation}\begin{split}
 \Lambda(r) = \left( \frac{r^2 \Phi_{rzz}}{
 3\Phi_{r}+r\Phi_{rr}-4r\Phi_{zz}} \right) (r,z=0).
 \end{split}\end{equation}
 They  show  that  $\Lambda(r)$ may  be   strongly linked to  the  mass
 gradient in the galactic plane.
 For the simplest cases we have that:

 1) for a spherical potential, $z_o=0$, $ \Lambda= 1 $ and the velocity
 ellipsoid points towards the galactic center,

 2) while for a cylindrical potential, $z_o=\infty$, $ \Lambda= 0 $ and
 the ellipsoid stays everywhere parallel to the galactic plane.


 \subsection{Measuring $z_o$}

 We    measure (\S4.1) $z_o$   directly from  Hipparcos  data using the
 distribution function Eq.~1.  Better constraints  are obtained using a
 sample with 3D velocities since $z_o$ is strongly tied to the coupling
 between $v_\theta$ and $v_z$ components  (\S4.2).

 Figure \ref{fig7} shows the  $f(v,w)$ marginal  velocity distribution
 for the 3D velocity  sample  and Fig.~\ref{fig8} the isocontours  for
 three models: the first model  is the best fit model  ($z_o=5.7\kpc$)
 and  two similar models just changing  $z_o$  to 0  and $\infty$.  We
 remark that these three models are strictly identical on the two axis
 $f(v,w=0)$ and  $f(v=v_{circ},w)$   (see   Appendix), and  that   the
 differences are important  for large velocities both  in $v$ and $w$.
 Outside these axis,   $f(v,w)$  depends quite exclusively  on  $z_o$.
 There is in fact a small dependence on $\alpha$, limited by the small
 range of acceptable $\alpha$ values from -.3 to -.1.  There is also a
 dependence  on the     kinematic scale  lengths   $R_{\sigma_r}$  and
 $R_{\sigma_z}$  that is  null when the  two scales  are equal  and is
 small when the scale lengths are large which is the actual situation.

 \subsection{Constraints on the galactic potential}

 Since the  determination ($z_o=5.7\pm1.4\kpc, \Lambda=0.69\pm.10$) is
 not yet very accurate,  we have not  developed detailed galactic mass
 models (uncertainties on  $R_{\sigma}$ and  $R_{\sigma_w}$ could lead
 at   most  to  $\Lambda=0.83$  corresponding    to  a  more spherical
 potential).  We mention here the  main conclusions that can be  drawn
 from the   ($z_o;\Lambda$)  measure in  the    galactic plane (longer
 developments are presented in Amendt \& Cuddeford, \cite{ac91}).

 For a   galaxy with a   flat rotation  curve and  a   full flat  mass
 distribution, $\Lambda(r)=0.5$  in  the    galactic plane.    For  an
 exponential  disc in a  flat rotation  curve  potential and inside  a
 spheroidal   halo,  $\Lambda(r)  \simeq   \frac{r} {4R_\Sigma}$  (the
 expression is  valid  at radii  where  the disc density  remains much
 larger than the   halo density).   For   a strictly   spherical  mass
 distribution  $\Lambda=1$.    Combining   components  with  different
 flattenings, a   large range of  values   from 0  to 1  and  more are
 possible.

 At  face  value,   ($z_o=5.7\kpc,   \Lambda=0.69$) implies  that   the
 potential distribution close to the sun is not extremely flattened and
 that not all the galactic mass is confined inside the galactic plane.

 If the galactic halo  is spherical, the  radial mass distribution  in
 the disc exponential and the rotation curve flat, then we deduce from
 the observed $\Lambda$ that the scale length of the mass distribution
 in  the disc is about  $2.7-3.1\kpc$ and is smaller  or larger if the
 rotation curve is respectively  decreasing or increasing.  Flattening
 the halo  is compensated by decreasing  the mass density scale length
 in the  disc  in  order to  keep   $\Lambda$ to the  observed  value.
 However the   dark matter halo  cannot  be flattened  by  more than a
 factor of 4 or   6, otherwise the  local   halo density is  not  much
 smaller than  the  disc mass   density and then  $\Lambda$  converges
 quickly towards 0.5.

 More  precise conclusions depend  on  adopted values  of the galactic
 structure such as the galactic solar radius $R_0$  and also the local
 stellar disc and  ISM mass density.  The  solar galactic radius $R_0$
 may be  smaller  than the IAU   recommended value (Reid, \cite{r93}).
 With  $R_0=7.5\kpc$ then $\Lambda=R_0/(4R_\Sigma)$  is closer  to the
 observed value.  A smaller  $R_0$ implies that  the rotation curve is
 locally    decreasing   (Rohlfs   \&  Kreitschmann,   \cite{rk88}) in
 concordance with our   findings from the  local velocity distribution
 (Tab.~2).

 \begin{sloppypar}
 When we model potentials with a  flat exponential disc and a Miyamoto
 halo  (with a core  radius and a flattening),   we do not find models
 with a  strictly   spherical   halo and having    sufficiently  small
 $\Lambda$ if the rotation curve is decreasing.   Halos with a density
 scale height  around $2\kpc$ at  $R_0$  give the  observed $\Lambda$.
 However this result depends critically  on the slope of the  rotation
 curve and  with flat or rising curve,  spherical halos (density scale
 height is $\sim 7\kpc$ at $R_0$) are obtained.
 \end{sloppypar}

 We conclude that an extremely flat dark halo is rejected at $2\sigma$
 and models favour a spherical halo or halos  with a local scale height
 larger  than $2 \kpc$ at  the solar galactic radius.   There is a need
 for more  data to  bring a stronger  constraint  and to determine more
 precisely the galactic potential.   This is also important because the
 basic measured quantities are  different from those determined in
 the  classical $K_z$ problem, and  consequently  this new method is an
 independent test  for measuring the mass  distribution in the galactic
 plane.

 We may reject with certainty some  galactic mass distributions: those
 corresponding to   (unexpected) oblate St\"ackel   potentials with  a
 focus beyond the solar galactic radius and an ellipsoid tilt pointing
 in the galactic plane outwards.

 Models  with $\Lambda=0$ are  also rejected: they could correspond to
 cylindrical potential  produced by a  vertical  and extremely prolate
 dark matter halo with a nearly constant mass  density up to height of
 about $10\kpc$.  Such  a  halo would  rotate about its  major axis in
 order to have an  angular momentum parallel  to the stellar disk one.
 This is a confirmation of collisionless simulations  of the dark halo
 that do not produce such halo configurations: ``There is decided lack
 of halos which rotate about the major axis  although this is a stable
 configuration'' (Warren et al, \cite{wq92}).

 \section{Conclusion}

 \begin{sloppypar}
 We analyse the  kinematics of nearby stars  from the Hipparcos proper
 motions and determine the solar motion and  the kinematic and density
 gradients of the observed stellar populations.   For that purpose, we
 have  extended a dynamical  model  (Shu, \cite{s69}) for axisymmetric
 exponential stellar discs to 3 dimensions.
 \end{sloppypar}

 We  determine the    solar motion relative  to   the   LSR  to  be  :
 $U_\odot=9.7\pm0.3\kms$,         $V_\odot=5.2\pm1.0\kms$          and
 $W_\odot=6.7\pm0.2\kms$.

 Assuming  $R_0=8.5\kpc$   and  $V_0=220\kms$, we determine  that  the
 stellar density scale length is short, $R_\Sigma=1.8\pm0.2\kpc$.

 The nonlinear shape of the  asymmetric drift is more likely explained
 assuming different $R_{\sigma_r}$ for young and old stars.  We find a
 large   kinematic    scale   length   for     blue   (young)    stars
 $R_{\sigma_r}=17\pm4\kpc$  and  for  red   stars  (predominantly old)
 $R_{\sigma_r}=9.7\pm0.8\kpc$ (or $R_{\sigma_r^2}=4.8\pm0.4\kpc$).

 These results are close to  previous kinematic determinations of  the
 scale lengths by  Mayor (\cite{m74}) ($R_\Sigma=2.2\kpc$) or Oblak \&
 Mayor           (\cite{om87})       ($R_{\sigma_r}=10\kpc$         or
 $R_{\sigma^2_r}=5\kpc$),    Fux    \&        Martinet   (\cite{fm94})
 ($R_\Sigma=2.5\kpc$), and Bienaym\'e \& S\'echaud (\cite{bs97}) based
 on the local stellar kinematics.

 We   also    constrain   locally    the rotation     curve.  Assuming
 $v_c(r)=r^\alpha$, we find $\alpha=-.24\pm0.04$.

 In the frame  of St\"ackel  potentials,  we explicitly  determine the
 dependence between vertical and horizontal motions.  We show that the
 marginal velocity distribution  of tangential and vertical velocities
 $f(v_\theta,v_z)$ allows us to constrain  the  shape of potential  in
 the solar neighbourhood through   the parameter $z_o$ (focus of   the
 coordinate system associated to   the fitted St\"ackel potential)  or
 $\Lambda$ (Eq.~13).

 From  3D-velocity      data,   we  measure   $z_o=5.7\pm1.4\kpc$   or
 $\Lambda=0.69\pm0.10$.  That quantity  is linked  to the total   mass
 gradient in the galactic plane.  It favours  a spherical dark halo or
 flattened halos  with a  local  thickness larger than  $2\kpc$.  More
 data would allow us to refine  this determination.  An extremely flat
 dark halo with all the mass in the galactic plane is ruled out.  This
 conclusion confirms the results obtained  measuring $K_z$, the  force
 perpendicular   to  the galactic plane,  and  the  local mass density
 (Cr\'ez\'e et al, \cite{cc98}), a   determination based on   vertical
 motions and  vertical density distribution   of tracer stars.   It is
 consistent with  preliminary results  from  Ibata et  al (\cite{i98})
 based on the  kinematics of the  stream associated to the Sagittarius
 dwarf galaxy through  the halo.   

 \appendix
 \section{disc distribution function in St\"ackel potentials}

 \subsection{St\"ackel potentials}

 \begin{sloppypar}
 A description   of 3D St\"ackel potentials  is   detailed in  de Zeeuw
 (\cite{dz85}).  St\"ackel  potentials  are more tractable  in confocal
 spheroidal coordinates.  The prolate spheroidal coordinates ($\lambda,
 \theta, \nu$) are related  to the cylindrical coordinates  ($r,\theta,
 z$) by
 \end{sloppypar}
 \begin{equation}
 r^2=\frac{(\lambda+\alpha)(\nu+\alpha)  }{ \alpha-\gamma}~~{\rm  and}~~~
 z^2=\frac{(\lambda+\gamma)(\nu+\gamma) }{ \gamma-\alpha}
 \end{equation}
 $\alpha$ and $\gamma$ determine the shape  of the coordinate surfaces,
 and   $\lambda,\,\nu$   satisfies $-\gamma\le\nu\le-\alpha\le\lambda$.
 Surfaces  of  constant $\lambda$  are spheroids  and those of constant
 $\nu$ are  hyperboloids.  They all share  the same foci located on the
 $z$ axis at $\pm z_o=\pm(\gamma-\alpha)^{1/2}$.  In the limit $z_o
 \rightarrow  \infty$, the coordinate  system  is cylindrical, and  for
 $z_o\rightarrow 0$ it is spherical.

 Then, the potential takes the form
 \begin{equation}
 \Phi(\lambda,\nu) = \frac{ h(\lambda)-h(\nu) }{ \lambda - \nu}
 \end{equation}
 where $h$ is an arbitrary  function ($h(\lambda)$ and $h(\nu)$ are not
 defined on the same interval).

 Properties of orbits  in such  potentials  are discussed by de   Zeeuw
 (\cite{dz85}).   They    all have  three   integrals  of  motions, for
 instance:
 \begin{equation}
 E = \Phi(\lambda,\nu)+\frac{1 }{ 2}(r\dot{\theta})^2 +\frac{1 }{ 2} \dot{r}^2
 +\frac{1 }{ 2} \dot{z}^2
 \end{equation}
 \begin{equation}
 L_z=r^2 \dot{\theta}=r v_\theta
 \end{equation}
 \begin{equation}\begin{split}
 I_3 = \Psi(\lambda,\nu)&-\frac{1 }{ 2}\frac{z^2}{ \gamma-\alpha}
 (\dot{r}^2+(r\dot{\theta)}^2) \\
 &-\frac{1 }{ 2}(\frac{r^2 }{ \gamma-\alpha}+1)\dot{z}^2
 + \frac{rz\dot{r}\dot{z} }{ \gamma-\alpha}
 \end{split}\end{equation}
 with
 \begin{equation}
 \Psi(\lambda,\nu)=\frac{ (\nu+\gamma)h(\lambda)-(\lambda+\gamma)h(\nu)
 }{  (\gamma-\alpha)(\lambda-\nu)}.
 \end{equation}

 $I_3$ is also the vertical energy  $E_z$ when the coordinate system is
 cylindrical, i.e. $z_o=\infty$.

 \subsection{3D disc distribution function}

 In a context similar  to this paper, i.e. ,  an analysis of a stellar
 velocity distribution, Statler (\cite{s89}) used St\"ackel potentials
 and built a distribution function valid towards the galactic pole. He
 showed that the change  of the orientation  of the velocity ellipsoid
 may bias  the estimated dynamical mass  in the disc  up to 10 percent
 for  samples at  a  distance of  $1  \kpc$ from  the  galactic plane.
 Analytic axisymmetric galaxy models  with  three integrals of  motion
 are also described by Dejonghe \& de Zeeuw (\cite{dd88}).

 Here, we give a 3D stellar disc distribution function that has nearly
 a  Schwarzschild distribution   behaviour  in  the  limit   of  small
 dispersions.  This  distribution has also a  density which  is nearly
 radially exponential.

 The distribution function is
 \begin{equation}\begin{split}
 f(E,L_z, I_3) =
 \frac{2\Omega(R_c)}{ 2\pi\kappa(R_c)}
 \frac{\Sigma(L_z)}{
            \sigma_r^2(L_z) }
 \exp \left[
 -\frac{E-E_{circ}}{\sigma_r^2}
 \right] \\
 \frac{ 1}{ \sqrt{2\pi} } \frac{1}{\sigma_z(L_z) }
 exp \left\{
 -(\frac{R_c(L_z)^2 }{ z_o^2 } +1 )^{-1}
 (\frac{1}{\sigma_z^2}-\frac{1}{\sigma_r^2})I_3
 \right\}
 \end{split}\end{equation}

 and depends on integrals of motion.

 This can be written differently as:
 \begin{equation}\begin{split}
 f(E,L_z,I_3)&=f_{3D}(r,z;\dot{r},\dot{\theta},\dot{z}) \\
 & = f_{//}(r;\dot{r},\dot{\theta})
 \frac{1}{ \sqrt{2\pi} \sigma_z}
 f_{\perp}(r,z;\dot{\theta},\dot{z})
 \end{split}\end{equation}
 with
 \begin{eqnarray}
 f_{//} (r;\dot{r},\dot{\theta})=
 \frac{2\Omega}{2\pi\kappa}
 \frac{\Sigma}{\sigma_r^2 }
 \exp \left[
 -\frac{E_{//}-E_{circ}}{\sigma_r^2}
 \right]
 \end{eqnarray}
 where   $E_{//}=\Phi+\frac{1}{2} (v_r^2+v_\theta^2)$  (when  z=0,  the
 expression  of $f_{//}$  is  exactly that   of $f_{2D}$,  i.e.  the 2D
 distribution function   in Bienaym\'e  \& S\'echaud, \cite{bs97})  and
 with
 \begin{equation}\begin{split}
 &f_{\perp}(r,z;\dot{\theta},\dot{z}) = \\
 &\exp \left\{  -\frac{1}{2}\frac{\dot{z}^2}{ \sigma_r^2}
 -(\frac{R_c(L_z)^2}{z_o^2} +1)^{-1}
 (\frac{1}{\sigma_z^2}-\frac{1}{\sigma_r^2})I_3 \right\}.
 \end{split}\end{equation}

 Properties  of $f_{3D}$ are  similar  to $f_{2D}$,  their  density and
 kinematics having nearly exponential radial decreases.

 The distribution function  simplifies when $z=0$, (i.e., $\nu=-\gamma$).
 Adding an  arbitrary constant to $h$ leaves  the potential unchanged, so
 without loss of generality, we may put $h(-\gamma)=0$, then:
 \begin{equation}
 \Phi(\lambda,\nu=-\gamma) =\frac {h(\lambda) }{ \lambda+\gamma }
 \end{equation}
 \begin{equation}\begin{split}
 \Psi(\lambda,\nu=-\gamma)=0
 \end{split}\end{equation}
 \begin{equation}
 I_3(z=0) = -\frac{1}{2}(\frac{r^2}{ z_o^2}+1)\dot{z}^2
 \end{equation}

 and finally we obtain:
 \begin{eqnarray}
 f_{\perp} =
 \exp \left\{
 -\frac{ \dot{z}^2}{ 2 \sigma_r^2}
 \left[ 1+(\frac{\sigma_r^2}{\sigma_z^2}-1)
 \left( \frac{ r^2+z_o^2 }{ R_c(L_z)^2 +z_o^2} \right) \right] \right\}
 \end{eqnarray}

 In   the case of  a  cylindrical  potential, radial  and perpendicular
 motions are uncoupled and $z_o=\infty$, so
 \begin{eqnarray}
 f_{\perp} =
 \exp \left\{ -\frac{\dot{z}^2}{ 2 \sigma_z^2} \right\}.
 \end{eqnarray}

 For a spherical  potential (i.e., $z_o =0$) and  a flat rotation curve
 $v_c(r)=v_o$, so $R_c=L_z/v_c$ and
 \begin{eqnarray}
 f_{\perp} =
 \exp \left\{
 -\frac{\dot{z}^2}{ 2}
 \left[ \frac{1}{\sigma_r^2}+(\frac{1}{\sigma_z^2}-\frac{1}{\sigma_r^2})
 \frac{ v_o^2 }{ v^2 }  \right] \right\}.
 \end{eqnarray}

 For a spherical  potential ($z_o =0$)  and a power  law rotation curve
 $v_c(r)=v_o(r/R_0)^\alpha$. If $r=R_0$, we have
 \begin{eqnarray}
 f_{\perp} =
 \exp \left\{
 -\frac{\dot{z}^2}{ 2}
 \left[ \frac{1}{\sigma_r^2}+(\frac{1}{\sigma_z^2}-\frac{1}{\sigma_r^2})
 \left(\frac{ v_o }{ v}\right)^{2/(1+\alpha)}   \right] \right\}.
 \end{eqnarray}
 \\
 Finally, we    remark the  two  following   simplifications  where the
 marginal distributions does not depend on $z_o$:

 -Firstly,  for  motions  in   and  parallel  to  the   galactic  plane
 ($v_z=0$, $z=0$ and $I_3=0$) we have
 \begin{equation}\begin{split}
 f_{3D}(r, z=0 ; v_r,v_\theta, & v_z=0)=\\
 & f_{//}(r;v_r,v_\theta)
 \frac{1}{ \sqrt{2\pi} \sigma_z},
 \end{split}\end{equation}

 and secondly when $z=0$ and $v_\theta=v_c(r)$, then $R_c=r$ and
 \begin{equation}\begin{split}
 f_{3D}& (r, z=0;v_r,v_\theta=v_c(r),v_z)= \\
 &f_{//}(r;v_r,v_\theta=v_c(r))
 \frac{1}{ \sqrt{2\pi} \sigma_z}
 \exp \left\{ -\frac{\dot{z}^2}{ 2 \sigma_z^2} \right\}.
 \end{split}\end{equation}

 \begin{sloppypar}
 It   is straightforward to integrate   (Eqs.~A18-19)  over the  radial
 velocities, then similar   expressions are obtained for  the  marginal
 velocity        distributions         $f(v_\theta,v_z=0)$          and
 $f(v_\theta=v_c(r),v_z)$  used in \S5.
 \end{sloppypar}

 \paragraph{Remark:}

 It is the requirement that the radial density  $(\int fd^3v)$ must be
 nearly   exponential    that   makes   one    introduce    the   term
 $(\frac{R_c(L_z)^2 }{ (\gamma-\alpha)}+1)$  in the second exponential
 of  Eq.~A7.  It  is  also  this  term that  introduces  the  peculiar
 dependence  in Eqs.~A16 or   A17  instead  of  having  the  uncoupled
 expression of Eq.~A15. If the density were different, the coupling in
 A16-17 would be different.


 \begin{acknowledgements}
 I would like  to thank  F. Ochsenbein, E.  Chereul,  C. Pichon and J. Pando
 for their help and comments.

 \end{acknowledgements}

 \end{document}